\documentclass{aa}  

\usepackage{graphicx}
\usepackage{txfonts}
\usepackage{xcolor}
\usepackage{threeparttable}
\usepackage{lscape}
\usepackage{rotating}

\bibpunct{(}{)}{;}{a}{}{,} % to follow the A&A style

\begin{document}

   \title{Characterisation of an EXor outburst SPICY 97589}

   %\subtitle{I. Overviewing the $\kappa$-mechanism}

   \author{A. Labdon
          \inst{1}
          \and
          R. A. B. Claes
          \inst{2}
          }

   \institute{European Southern Observatory, Casilla 19001, Santiago 19, Chile \\
              \email{alabdon@eso.org}\\
              European Southern Observatory, Karl-Schwarzschild-Strasse 2, 85748 Garching bei München, Germany
              \
             }

   \date{Received September 15, 1996; accepted March 16, 1997}

% \abstract{}{}{}{}{} 
% 5 {} token are mandatory
 
  \abstract
   {Stellar outbursts from variable or periodic accretion are thought to be ubiquitous across young stellar populations. However, relatively few outbursting objects have been discovered to date. Here, we present the characterisation of a new EXor-type episodic accretor.}
   {We aim to characterise the nature of the 2023 outburst of SPICY 97589/Gaia23bab and characterise the stellar source for the first time, while exploring how an accretion outburst contributes to disk evolution.}
   {We employ multi-waveband medium-resolution spectroscopy with UVB-VIS-NIR coverage during the peak of the 2023 outburst and the post-outburst quiescent object. The broad wavelength coverage of the dataset allows for robust measurements of the accretion rate using known line tracers. The addition of quiescent spectra provides a good estimation of stellar parameters of the central star while also informing us on the evolution of the disk during outburst phases. }
   {We find the stellar source to be a 3410\,K, M3.0 type star with a luminosity of 0.41 $L_\odot$ and an estimated stellar mass of 0.29 $M_\odot$. We measure the accretion rate of SPICY 97589 to be $\dot M = 2.38\pm0.58\times10^{-7}\,\mathrm{M_\odot yr^{-1}}$. This value is at two orders of magnitude greater than the quiescent accretion rate. Thus, we confirm that the 2023 outburst was driven by an influx of material from the surrounding environment to the central star, an accretion outburst. The spectral fingerprint of emission lines is also characteristic of an outbursting EXor-type source, including variable disk winds.}
   {}

\keywords{Stars: variables: T Tauri, Herbig Ae/Be – Techniques: spectroscopic – Protoplanetary disks
   }

   \maketitle
%
%-------------------------------------------------------------------

\section{Introduction}

   Accretion onto astronomical objects is one of the most fundamental processes in astrophysics. It facilitates mass transport onto a wide range of astrophysical objects, from planets and stars to super-massive black holes \citep{Lin96}. The mass transport proceeds through accretion disks, where viscosity converts angular momentum into thermal energy, thus enabling the mass infall \citep{Pringle72}.

   Active accretion disks have been observed around a wide range of YSO (Young Stellar Object) classes. However, YSOs are known to be 10-100 times less luminous than expected from steady-state accretion scenarios \citep{Dunham12}. Particularly given the typical accretion rates of the order $10^{-9}\,\mathrm{M_\odot yr^{-1}}$ observed around many YSOs. This raises the possibility that accretion is not a steady-state process across the early stages of stellar evolution, but is episodic \citep{Kenyon95,Evans09,Fischer23}. Episodic accretion can occur on various scales, ranging from low-level variability to dramatic outbursts.

   Accretion outbursts are difficult to classify into distinct groups; as individual occurrences vary greatly in strength, duration, onset and cause. However, they are broadly split into two types, FUor and EXor \citep{Fischer23} based on the duration and intensity of the outburst. FUors (named for archetype FU\,Orionis) are characterised by rapid brightening events of $4-6\,\mathrm{mag}$ \citep{Audard14} followed by a protracted period of dimming, on the order of decades or centuries \citep{Hartmann96,Herbig07,kra16}. On the other hand, EXors (named for archetype EX\,Lupi), are characterised by shorter outbursts on the order of months, with smaller peaks of $2-3\,\mathrm{mag}$. The rise and fall of the light curve to and from quiescence are roughly equal in duration and can typically recur every few years/decades \citep{Audard14}. It has become increasingly accepted that most YSOs will exhibit some form of extreme episodic accretion, throughout their lifetime \citep{Hartmann96,Audard14}.

   On 2023-03-06, a Gaia photometric alert was issued for the YSO SPICY 97589, dubbed Gaia23bab \citep{Hodgkin23}. At this stage, the object had already been brightening for over one year and was already at peak outburst. Shortly following the alert, an initial characterisation was completed by \citet{Kuhn23}. They determined the basic characteristics of the visible light curves, in addition to distances and extinction measurements, based on the SED and Gaia parallax measurements. However, as they note, spectroscopic follow-up was required to characterise the source fully. 

   \citet{Giannini2024} produced an initial characterisation of SPICY 97589 using a combination of photometric light curves and NIR spectroscopic observations taken around May/July 2023, around 2/3 months after the peak of the outburst. Based on the SED, they characterised the central star as G3-K0 type with an effective temperature of 5400\,K and a mass of $\mathrm{1.6\pm0.1}\,M_\odot$. However, this approach is limited due to strong, likely variable extinction in the SED and the broad range of assumptions required. From the NIR spectra \citet{Giannini2024}, they were able to estimate the accretion rate at the time to $\dot M = 2.0\pm1.0\times10^{-7}\,\mathrm{M_\odot yr^{-1}}$ using similar methods employed in this paper, which provides an important comparison.

   \citet{Nagy25} continued the analysis with a detailed study using optical and NIR spectra SPICY 97589. Based on analysis of accretion tracing hydrogen lines, they measured the accretion rate at $\dot M \sim 2.0\pm1.0\times10^{-7}\,\mathrm{M_\odot yr^{-1}}$, corroborating the work of \citet{Giannini2024}. They also performed modelling of the Balmer, Paschen, and Calcium lines to estimate the hydrogen density and excitation temperatures of the outburst. 

%These previous studies have lacked quiescent spectra for direct comparison to
   
    \begin{table}[t]
        \centering
        \begin{tabular}{c c} 
            \hline
            \noalign{\smallskip}
            Parameter  &  Value   \\ [0.5ex]
            \hline
            \noalign{\smallskip}
            UVB Slit [arcsec] & 0.8x11 \\
            %\noalign{\smallskip}
            UVB EXP [s] & 1500 \\
            %\noalign{\smallskip}
            UVB NEXP & 2*  \\
            %\noalign{\smallskip}
            UVB readout mode & high gain \\
             & 1x2 binning  \\
            %\noalign{\smallskip}

            \hline
            \noalign{\smallskip}
            VIS Slit [arcsec] & 0.7x11\\
            %\noalign{\smallskip}
            VIS EXP [s] & 1450 \\
            %\noalign{\smallskip}
            VIS NEXP & 2*  \\
            %\noalign{\smallskip}
            VIS readout mode & high gain \\
             & 1x2 binning  \\
            %\noalign{\smallskip}

            \hline
            \noalign{\smallskip}
            NIR Slit [arcsec] & 0.6x11\\
            %\noalign{\smallskip}
            NIR DIT [s] & 1200 \\
            %\noalign{\smallskip}
            NIR NDIT & 2*  \\
            %\noalign{\smallskip}
            
            \hline
            \noalign{\smallskip}
        \end{tabular}
        \caption{\label{table:XSHOOTER} Parameters of the X-Shooter observations of SPICY 97589, taken on 2023-04-19. EXP is the exposure time, NEXP is the number of exposures. DIT is the integration time (for NIR detectors), and NDIT is the number of integrations. NDIT and NEXP were increased to 3 on all arms for the follow-up observations during quiescence. Other parameters remained the same.}
    \end{table}

   This paper presents new spectroscopic follow-up to an EXor-type outbursting YSO using multi-waveband UVB-VIS-NIR medium-resolution spectrometry from the X-Shooter instrument. In section\,\ref{sec:Observations}, we detail the origin of the data and the reduction techniques we employed. In section\,\ref{sec:Results}, we present the data and provide simple descriptions. In section\,\ref{sec:Discussion}, we analyse the results in detail and provide context with comparisons to other outbursting sources. Finally, in section\,\ref{sec:Conclusion}, we offer the concluding statements.

%--------------------------------------------------------------------
\section{Observations} \label{sec:Observations}

    \subsection{Light Curves}

    Photometric monitoring is the only reliable method to detect outbursting objects; a high cadence across wide parts of the sky is required to find such events. The outbursting nature of the source was first identified using the Gaia satellite. Gaia photometry covers a wide G band from $330\,\mathrm{nm}$ to $1050\,\mathrm{nm}$, from near-ultraviolet to near-infrared. The cadence of GAIA photometry varied greatly, with clusters of observations sometimes separated by a few months. 

    In addition, mid-infrared (MIR) light curves were available from the WISE satellite. The light curves are compiled using a Python package \footnote{https://github.com/HC-Hwang/wise\_light\_curves} to access the data and filter bad photometric measurements \citep{Hwang20}. SPICY 97589 has been observed with a low cadence ($4-6$ months) in the W1 ($3.35\,\mathrm{\mu m}$) and W2 ($4.6\,\mathrm{\mu m}$) bands. Although the cadence differs from Gaia's, the light curves clearly show the two outbursts and provide important colour information about the outbursts. 

    \subsection{Spectroscopic Data}

    \begin{figure}[b]
        \centering
        \includegraphics[scale=0.55]{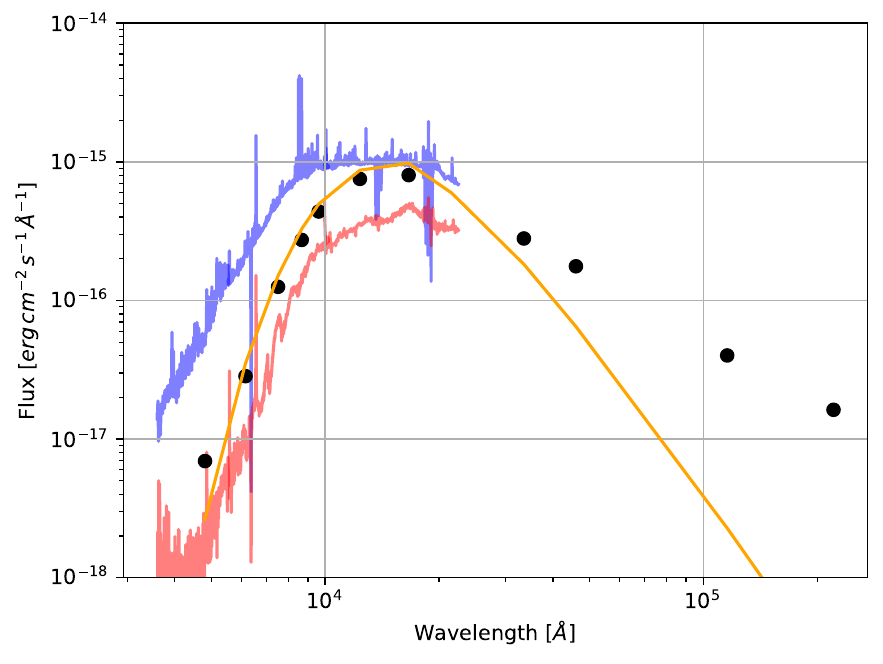}
        \caption{Spectral energy distribution (SED) of SPICY 97589. Circles are the photometry from sources described in Appendix\,\ref{AppA}. All photometry is taken during quiescence, pre-2017 outburst. The blue line is the binned X-Shooter flux calibrated 2023 outburst spectrum. The red line is the binned X-Shooter flux calibrated spectrum post-2023 outburst.} %The red circles represent the closest photometric measurements to the peak of the outburst from Gaia ($\sim 0.7\,\mathrm{\mu m}$, and NEOWISE ($3.5$ and $4.6\,\mathrm{\mu m})$, The NEOWISE photometry was taken $\sim$ 2 months before the peak of the outburst.}
        \label{fig:SED}
    \end{figure}

    Shortly after issuing the GAIA photometric alert Gaia23bab, we obtained spectroscopic observations with the X-Shooter \citep{Vernet11} instrument from the European Southern Observatory (ESO). The time was obtained through the Directors Discretionary Time (DDT) program 111.263U. X-Shooter is an intermediate-resolution echelle spectrograph operating from $0.3$ to $2500\,\mathrm{\mu m}$. Such wide wavelength coverage is ideal for observing accreting objects, providing access to crucial spectral lines in the visible (VIS) and near-infrared (NIR) and the accretion-related Balmer jump in the ultra-violet (UVB).

    Observations were obtained on 2023-04-19, only two months after the peak of the outburst, at which point the brightness was still $\sim 2$ magnitudes above quiescence. The parameters of the observations are shown in Table\ref{table:XSHOOTER}. The slit widths of 0.8", 0.7" and 0.6" of the UVB, VIS and NIR arms were chosen to provide the maximum spectral resolution while maintaining reasonable integration times on the object. The spectral resolution of the arms is $R = 6700$, $11400$ and $8100$ for the UVB, VIS and NIR arms, respectively. 

    Follow-up observations were obtained on 2024-07-25 after the object had returned to its quiescent state once again through DDT program 112.26Z7. Care was taken to ensure the same instrument setup was used but with increased total exposure time to account for the reduced quiescent brightness of the object. During the quiescent period, the signal-to-noise ratio (SNR) of the UVB arm was very low, even with the increased total exposure time. 
    
    All data were reduced using the standard X-Shooter data reduction pipeline (v. 3.6.3) \citep{Modigliani10}; the pipeline utilises a spectro-photometric standard star to provide absolute flux calibrations across all three arms. A telluric correction was then applied using Molecfit (v.1.5.9) \citep{Smette15}. Molecfit uses a telluric standard star, observed in the same setup and on the same night as the science observations, to calculate the atmospheric parameters and create a model telluric spectrum for subtraction from the science data. The result is absolute flux calibrated, telluric subtracted spectra from 300 to 2400\,nm. The exact same procedure was followed for both the outburst and quiescent epochs to ensure consistent analysis. 

    \begin{figure}[b]
    \centering
    \includegraphics[width=0.99\linewidth]{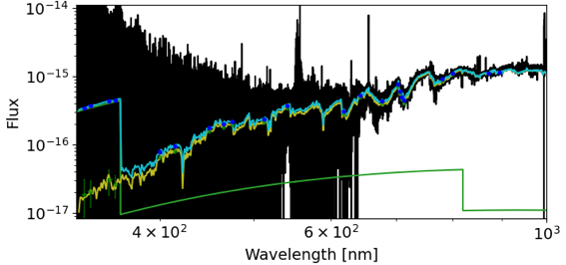}
    \caption{Best fit of the X-Shooter spectrum of SPICY 97589. The observed spectrum is shown in black. The best fit photospheric template in yellow, and the slab model in green. The best fit is shown in light blue. The blue points indicate the wavelength ranges used to constrain the model.}
    \label{fig:bestFit}
    \end{figure}

    \begin{figure*}[t]
    \centering
    \includegraphics[scale=0.6]{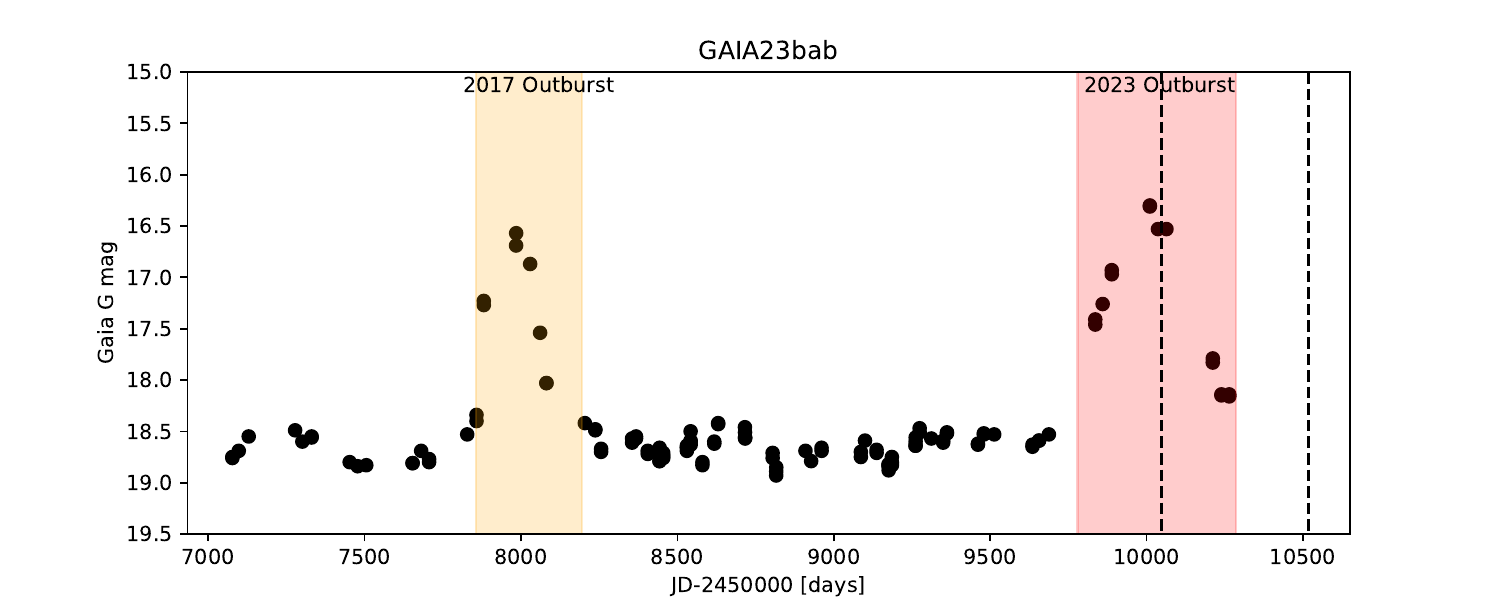}
    \caption{Gaia G-band light curve over the past seven years. Highlighted in orange and red are the 2017 and 2023 outbursts, respectively. The black dashed lines are the data of the X-Shooter spectroscopic observations on 2023-04-19 and 2024-07-25. The mean quiescent magnitude is measured as $18.66\pm0.12$.}
    \label{fig:GAIAlightcurve}
    \end{figure*}

\section{Results} \label{sec:Results}

\subsection{The Underlying Star} \label{Stellar}

To obtain the stellar properties of SPICY 97589 we use the FRAPPE tool presented by \citet{Claes2024}. The photospheric features in the outburst spectrum are too veiled to accurately constrain the stellar properties. The entire spectrum is dominated by broad emission lines/bands, while FRAPPE only includes continuum hydrogen emission, representing an accretion shock not disk emission, which is likely more complex. Therefore, we applied FRAPPE only to the post-outburst spectrum. FRAPPE fits X-Shooter spectra using a grid of accretion slab models to represent the emission originating from the accretion shock, an interpolated Class III template to represent the stellar photosphere and an extinction law. In addition to the wavelength ranges used in \citet{Claes2024}, we use additional wavelength ranges at $\sim 619$ nm, $\sim 644$ nm, $\sim 647$ nm, $\sim 781$ nm, $\sim 801$ nm, $\sim 871$ nm, and $\sim 895$ nm in the best fit determination to provide a better constraint on the stellar parameters. Veiling is accounted for by including the accretion slab model simultaneously with the extinction and spectral type, and is modelled using the depth of several bands, including TiO absorption at $\sim714\,\mathrm{nm}$.

The best fit of the post-outburst epoch is shown in Figure \ref{fig:bestFit}. The best-fitting model has spectral type M3.0, corresponding to an effective temperature of 3410 K and an extinction of $A_V$ = 4.4, using the \citet{cardelli98} extinction law with $R_V = 3.1$. Adopting a distance of 900 PC \citep{Giannini2024}, we find a stellar luminosity of $0.41\pm0.09\,L_\odot$. Using the isochrones of \citet{B15}, we estimate a stellar mass of $0.29\pm0.05\,M_\odot$ and a stellar age of $0.75\pm0.5\,Myr$, although the uncertainty in isochronal ages for individual targets can be significant. Although it should be noted that individual ages can be unreliable for individual targets.%\rc{Isochronal ages tend to be unreliable for individual target, not sure if you want to include this} 

The signal-to-noise at wavelengths around and shorter than the Balmer jump was too low to constrain the accretion slab model in this region. Since the majority of the accretion emission occurs at these wavelengths, we cannot use the slab model to provide an accurate estimate of the accretion luminosity. Therefore, we do not report the accretion properties associated with the best fit.

\subsection{Light Curves}

    \begin{figure*}[]
    \centering
    \includegraphics[scale=0.6]{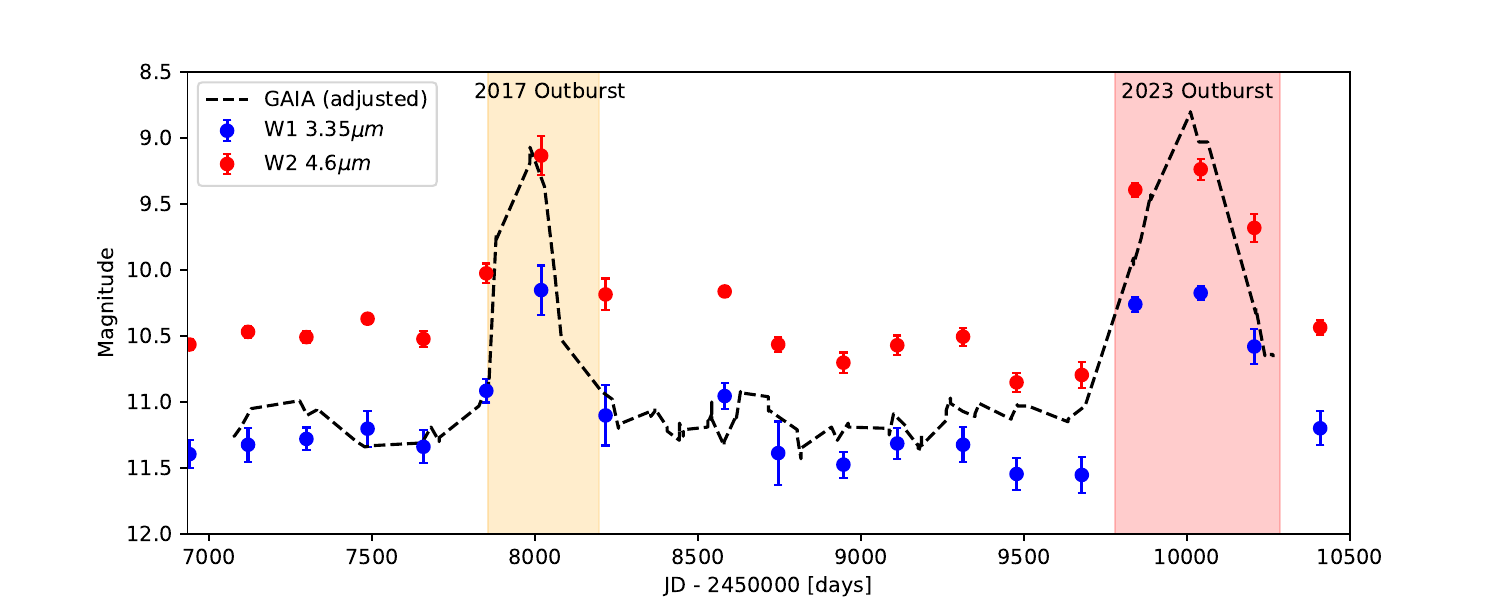}
    \caption{WISE light curves in the W1 ($3.35\,\mathrm{\mu m}$) and W2 ($4.6\,\mathrm{\mu m}$) shown in blue and red respectively. For each date of WISE observations, multiple images are taken, and the photometry of each image is averaged to provide the final values. The black dashed line represents the Gaia G band light curve, adjusted in magnitude scale by 7.5mag. Highlighted in orange and red are the 2017 and 2023 outbursts, respectively.}
    \label{fig:WISElightcurve}
    \end{figure*}

The Gaia light curves are shown in Figure\,\ref{fig:GAIAlightcurve}. Two outbursts are visible, the first in 2017 and the second in 2023. By excluding the outburst ranges, we can measure the mean magnitude of the star during quiescence as $18.66\pm0.12$ magnitudes in the G band. The parameters of the outbursts can be measured by fitting simple models to the outburst light curves, which approximately follow Gaussian distributions.% A summary of the results is provided in Table\,\ref{table:OutburstFits}

In 2017, the peak outburst was 2.24 mag above the average quiescence magnitude. The FWHM of the curve is 172 days, and the total duration is 358 days. We define the duration and total time the fitted Gaussian is above the upper limit of the quiescence magnitude, thus providing a lower limit to the actual duration. For 2023, the peak outburst is 2.51 mag above the mean baseline. The FWHM of the outburst is 237 days, and the total duration is 671 days. In summary, the 2023 outburst was slightly stronger and significantly more extended in duration than the 2017 outburst. 

The WISE light curves are more limited due to the lower cadence of observations, but which nevertheless cover the full 2017 and 2023 outbursts. In both outbursts, a pointing was obtained very close to the peak of the outburst. For the W1 band at $3.35\,\mathrm{\mu m}$, we measure the quiescence baseline at $11.29\pm0.19$ mag, with the peak outburst magnitude at $1.15\pm0.31$ above the baseline in 2017 and $1.17\pm0.24$ above the baseline in 2023. For the W2 band at $4.6\,\mathrm{\mu m}$, we measure the quiescence at $10.49\pm0.23$ mag, with the peak outburst magnitude at $1.36\pm0.38$ above the baseline in 2017 and $1.25\pm0.30$ above the baseline in 2023. Overall, the magnitude of both outbursts is greater at visible wavelengths than at mid-infrared wavelengths probed by WISE.

\subsection{Measuring Accretion Rates}
\begin{landscape}
\begin{figure}
    \includegraphics[width=1.3\textwidth]{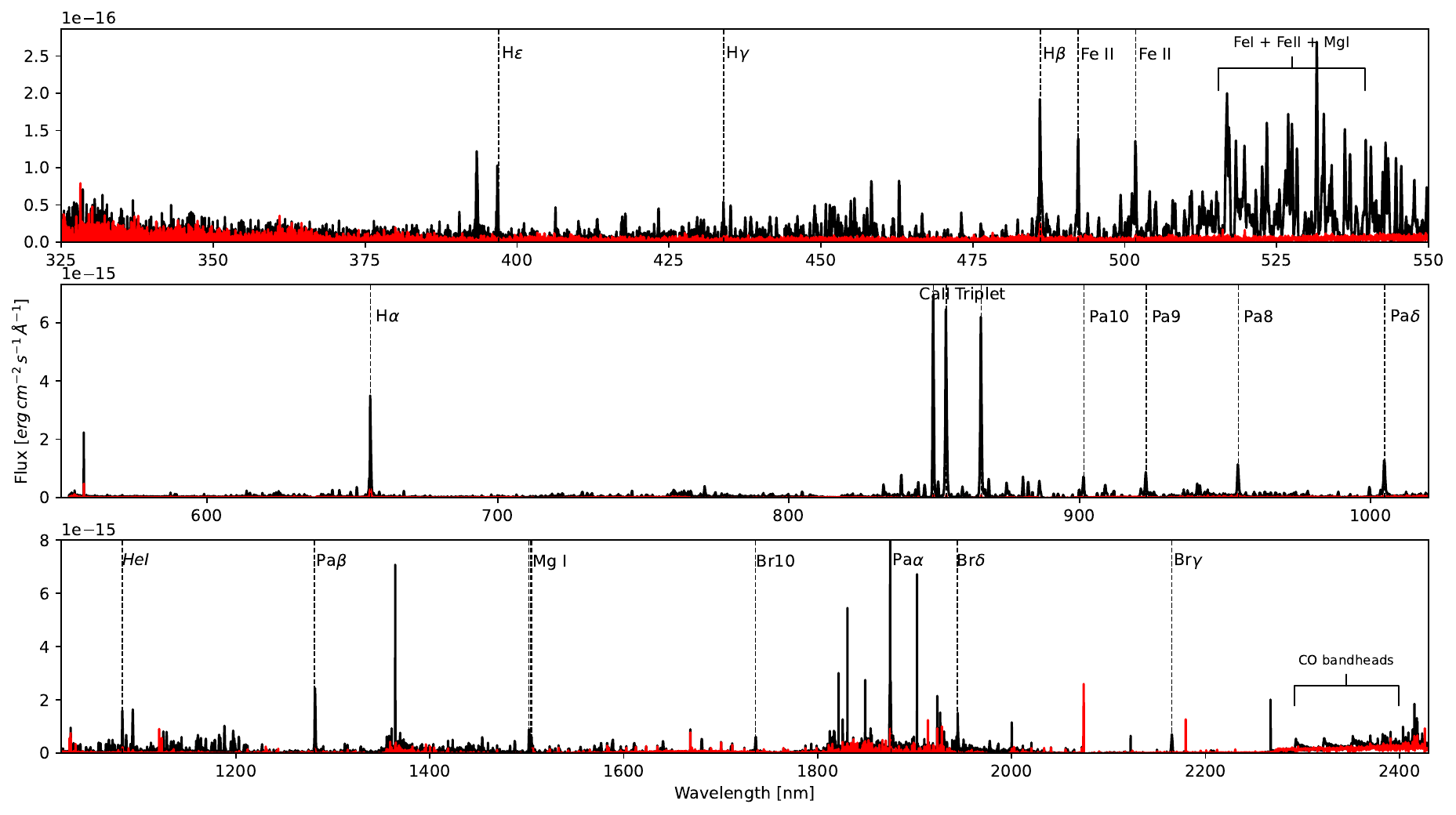}
    \caption{Continuum-subtracted X-Shooter spectra across the three instrument arms. The three panels from top to bottom show the spectra from the UVB, VIS and NIR arms. In black is the outburst spectra and in red is the quiescent spectra. Highlighted as dashed lines are the prominent emission lines present in the spectra; a full list of the identified lines is presented in Appendix A.}
    \label{fig:AllSpectra}
\end{figure}
\end{landscape}

    The measurement of accretion rates from spectral lines is possible due to the intrinsic relation between hydrogen emission and accretion. Most commonly used are the H$\mathrm{\alpha}$ ($656.3\,\mathrm{nm}$), H$\mathrm{\beta}$ ($486.1\,\mathrm{nm}$), Pa$\mathrm{\beta}$ ($1282.4\,\mathrm{nm}$) and Br$\mathrm{\gamma}$ ($2166.5\,\mathrm{nm}$) lines, all of which strongly feature in SPICY\,97589. A full list of the lines used to measure the mass accretion rate is shown in Table\,\ref{table:AccretionRates}, with many also highlighted in Figure\,\ref{fig:AllSpectra}. Overall, 22 lines are identified that correlate with the accretion rate, mostly hydrogen lines but also Helium-I and Calcium-II lines. The Balmer series lines are unreliable in the outbursting spectra due to strong P-Cygni profiles impacting the equivalent width and line luminosity measurements. We follow the example of \citet{Alcala14}, \citet{Fairlamb17}, and later \citet{Vioque22} to equate the equivalent width of the spectral lines to the mass accretion rate. The equivalent widths are measured from the de-reddened and continuum-subtracted spectra, using the extinction values calculated in Section\,\ref{Stellar} and following the example of \citet{Fitzpatrick99}.

    It should be noted that changes in extinction during episodic accretion events are possible, as shown in \citet{Lorenzetti12}. In particular, extinction decreases as obscuring material can be cleared away due to enhanced winds and outflows. This can cause an underestimation of the accretion rate. Lower values of extinction increase the equivalent width of the line tracers, increasing the accretion luminosity and inferred accretion rate. The values we compute below are, therefore, conservative in their estimation of mass accretion. 
    
    To covert line widths to mass accretion rate, we use the equivalent width ($EW$) of the relevant lines to obtain the line flux ($F_{H\,line}$) though $F_{H\,line} = EW \cdot F_\lambda$ where $F_\lambda$ is the estimated continuum flux at the central wavelength of the line. The line luminosity can then be calculated using:
    \begin{equation}
        log(L_{acc} / L_\odot) = A + B \cdot log(L_{H\alpha,\beta}/L_\odot),
    \end{equation}
    $L_{acc}$ and $L_\odot$ are the accretion and solar luminosities, respectively, and $A$ and $B$ are constants. For T\,Tauri stars, \citet{Alcala17} compiles a list of constants for the common accretion lines, for example $A = 1.13\pm0.05$ and $B = 1.74\pm0.19$ for $H_\alpha$. Various additional Brackett transitions are also compiled in \citet{Fairlamb17} and are included here. The values for all the lines used for measuring accretion rates can be found within \citet{Alcala17}. The higher order Brackett transitions are described in \citet{Fairlamb17} as being less sensitive to accretion due to decreasing strength in emission at shorter wavelengths. These Brackett transitions are observed only in outburst spectra and are absent in quiescent spectra.

     From the accretion luminosity, the mass accretion rate can be derived as:
    \begin{equation}
        \dot M_{acc} = \frac{L_{acc}R_*}{GM_*} = \frac{L_{acc}}{GM_*} \cdot \sqrt{ \frac{L} {4\pi\sigma T^4_{eff} }  } ,
    \end{equation}
    where $\dot M_{acc} $ is the mass accretion rate, $R_*$ is the stellar radius, $M_*$ is the stellar mass, $G$ is the gravitational constant and $T_{eff}$ is the effective temperature.

    The measured line parameters and derived accretion rates for each line are shown in Table\,\ref{table:AccretionRates}. The final calculated accretion rate during the outburst is $2.38\pm0.58\times10^{-7}$, and during quiescence is $4.24\pm0.58\times10^{-9}$. At the peak of the outburst, the accretion is two orders of magnitude greater than post-outburst quiescence. The errors are compounded from many sources: the initial values of $A$ and $B$, uncertainties in the underlying stellar model computed from FRAPPE, and uncertainties in the equivalent width measurements arising from instrumental errors in the original X-Shooter spectra.

    An alternative method for measuring the accretion rate in young stars is to measure the magnitude of the Balmer jump in the $300-400\,\mathrm{nm}$ range of the UV spectrum \citep{Herczeg08}. However, the signal-to-noise of the lower UV spectrum in our data is low, and there is no significant evidence of excess across the Balmer jump, so this method cannot be used accurately.

    %The accretion rates measured from $H_{\alpha}$, $Pa_{\beta}$ and $Br_{\gamma}$ are all in good agreement, given the limitations of the method, with an average mass accretion rate of $\dot M = 6.8\pm1.8\times10^{-7}\,\mathrm{M_\odot yr^{-1}}$. However, $H_{\beta}$ finds a significantly high accretion rate; this may be due to the underlying noise of the UVB spectra impacting the equivalent width measurements. As this value is $\geq5\,\sigma$ different from the other values, we declare this result anomalous. 

\begin{table*}
\centering
\begin{threeparttable}
    \caption{\label{table:AccretionRates} Summary of the line parameters of the 4 common accretion tracers explored in this work, along with the calculated accretion rates. EW is the line equivalent width, $L_{acc}$ is the associated accretion luminosity, and $\dot M$ is the calculated accretion rate. } 
    \centering
    \begin{tabular}{c c c c c} 
        \hline
        \noalign{\smallskip}
        Line Tracer  &  $\lambda$ [nm] &  $L_{acc}$ Outburst & $\dot M$ Outburst $[\mathrm{M_\odot yr^{-1}}]$ & $\dot M$ Quiescent $[\mathrm{M_\odot yr^{-1}}]$\\ [0.5ex]
        \hline
        $\mathrm{H\alpha}$ &	656.3	&	$3.51\pm3.24\times10^{-4}$	&	$4.96\pm1.94\times10^{-10}$* &	$5.32\pm2.37\times10^{-11}	$\\
        \noalign{\smallskip}
        $\mathrm{H\beta}$	&	486.1	&	$7.39\pm0.88\times10^{-6}$	&	$1.05\pm1.30\times10^{-11}$* & ** \\
        \noalign{\smallskip}
        $\mathrm{H\gamma}$	&	434.1	&	$5.66\pm1.26\times10^{-8}$	&	$8.01\pm1.34\times10^{-14}$* & ** \\
        \noalign{\smallskip}
        $\mathrm{Pa\beta}$	&	1281.8	&	$1.98\pm0.56\times10^{-1}$	&	$2.80\pm0.85\times10^{-7}$ &	$6.29\pm1.81\times10^{-9}	$ \\
        \noalign{\smallskip}
        $\mathrm{Pa\gamma}$	&	1093.8	&	$6.55\pm0.19\times10^{-2}$	&	$9.27\pm0.0.61\times10^{-8}$ &	$2.50\pm1.81\times10^{-9}	$\\
        \noalign{\smallskip}
        $\mathrm{Pa\delta}$	&	1004.9	&	$8.73\pm0.29\times10^{-2}$	&	$1.23\pm1.25\times10^{-7}$ &	$7.87\pm1.61\times10^{-10}	$\\
        \noalign{\smallskip}
        $\mathrm{Pa8}$	&	954.6	&	$5.03\pm0.15\times10^{-2}$	&	$7.12\pm0.38\times10^{-8}$ &	$2.26\pm1.59\times10^{-9}	$\\
        \noalign{\smallskip}
        $\mathrm{Pa9}$	&	922.9	&	$5.66\pm0.13\times10^{-2}$	&	$8.00\pm0.42\times10^{-8}$ &	$6.38\pm1.48\times10^{-11}	$\\
        \noalign{\smallskip}
        $\mathrm{Pa10}$	&	901.5	&	$3.96\pm0.13\times10^{-2}$	&	$5.60\pm0.44\times10^{-8}$ &	$1.43\pm0.94\times10^{-10}	$\\
        \noalign{\smallskip}
        $\mathrm{Br\gamma}$	&	2165.3	&	$1.02\pm0.03$	&	$1.44\pm0.19\times10^{-6}$ &	$3.57\pm2.51\times10^{-8}	$\\
        \noalign{\smallskip}
        $\mathrm{Br(11-4)}$	&	1680.6	&	$3.78\pm0.03\times10^{-2}$	&	$3.73\pm0.25\times10^{-7}$ & ** \\
        \noalign{\smallskip}
        $\mathrm{Br(12-4)}$	&	1640.7	&	$3.26\pm0.02\times10^{-2}$	&	$3.25\pm0.28\times10^{-7}$ & ** \\
        \noalign{\smallskip}
        $\mathrm{Br(13-4)}$	&	1610.9	&	$5.34\pm0.04\times10^{-2}$	&	$5.29\pm0.35\times10^{-7}$ & ** \\
        \noalign{\smallskip}
        $\mathrm{Br(14-4)}$	&	1588.0	&	$5.70\pm0.06\times10^{-2}$	&	$3.43\pm0.24\times10^{-7}$ & ** \\
        \noalign{\smallskip}
        $\mathrm{Br(15-4)}$	&	1570.1	&	$5.61\pm0.05\times10^{-2}$	&	$5.55\pm0.42\times10^{-7}$ & ** \\
        \noalign{\smallskip}
        $\mathrm{Br(16-4)}$	&	1555.6	&	$5.56\pm0.04\times10^{-2}$	&	$3.81\pm0.29\times10^{-7}$ & ** \\
        \noalign{\smallskip}
        $\mathrm{Br(17-4)}$	&	1543.9	&	$5.40\pm0.05\times10^{-2}$	&	$3.47\pm0.25\times10^{-7}$ & ** \\
        \noalign{\smallskip}
        $\mathrm{He I}$	&	1082.9	&	$3.31\pm0.40\times10^{-2}$	&	$4.69\pm0.38\times10^{-8}$ &	$1.24\pm1.15\times10^{-10}	$\\
        \noalign{\smallskip}
        $\mathrm{He I}$	&	587.6	&	$1.21\pm1.47\times10^{-4}$	&	$1.71\pm1.16\times10^{-10}$ &	$2.92\pm2.28\times10^{-12}	$\\
        \noalign{\smallskip}
        $\mathrm{Ca II}$	&	849.8	&	$4.29\pm0.55\times10^{-2}$	&	$6.07\pm0.61\times10^{-8}$ &	$7.41\pm1.01\times10^{-10}	$\\
        \noalign{\smallskip}
        $\mathrm{Ca II}$	&	854.2	&	$6.97\pm0.14\times10^{-2}$	&	$9.87\pm0.78\times10^{-8}$ &	$2.83\pm1.65\times10^{-10}	$\\
        \noalign{\smallskip}
        $\mathrm{Ca II}$	&	866.2	&	$1.09\pm0.18\times10^{-1}$	&	$1.54\pm0.41\times10^{-7}$ &	$1.28\pm0.87\times10^{-9}	$\\
        \noalign{\smallskip}
        \hline
        \noalign{\smallskip}
    \end{tabular}
    \begin{tablenotes}
      \small
      \item * Measurements are strongly impacted by p-cyngi profiles in the line; as such, they are not considered when computing the final accretion rate. ** Line too faint for accurate measurements
    \end{tablenotes}
  \end{threeparttable}
\end{table*}

\subsection{Hydrogen Emission Lines}

A forest of hydrogen emission lines across the spectra covers the Balmer, Paschen and Brackett series, mostly only detected in the outbursting spectrum. Appendix\,\ref{AppB} contains a complete list of the observed hydrogen lines. The Paschen and Brackett lines are in strong emission, while all the Balmer lines exhibit clear P-Cygni profiles, with blue-shifted absorption and red-shifted emission. The P-Cygni profile of the Ba$\alpha$ line at $656.3\,\mathrm{nm}$ is shown in Figure\,\ref{fig:BaAlpha}. The Balmer lines change dramatically between the peak and post-outburst spectra. During the outburst, very strong emission and P-Cygni profiles dominate. During the quiescent phase, a fainter double-peaked emission is seen, without a hint of P-Cygni absorption. This indicates a different origin for these lines, likely due to a distinct physical process. The other hydrogen lines do not show this feature; they just change in emission intensity. The origin and interpretation of the Hydrogen lines is discussed further in Section\,\ref{sec:SpectraDis}.

\section{Discussion} \label{sec:Discussion}

These observations paint an intriguing picture of the evolution of a YSO undergoing outburst. By combining peak and post-outburst data, we have been able to characterise the underlying star, explore the nature of the outburst and comment on the post-outburst effects on the circumstellar environment.

\subsection{Stellar Classification}

Our stellar classification using the FRAPPE tool found an M3.0 type star with an effective temperature of 3410\,K and a strong extinction profile of $A_V = 4.4$. This is similar to the work of \citet{Kuhn23} who estimated a temperature of 4700\,K at $A_V \sim 5.0$. However, this work used only outburst spectra and pre-outburst photometry to make these estimates.

We take this further through the use of isochrone fitting to provide a broad estimate of the mass and age of the system at $0.29\pm0.05\,\mathrm{M_\odot}$ and $0.75\pm0.5\,\mathrm{Myr}$, respectively. The spectral type and mass are broadly representative of other outbursting YSOs, EX\,Lupi has a mass of $\sim 0.6\,M_\odot$ \citep{Sipos09}, and V1118\,Ori another typical EXor has a mass of $0.41\pm0.09\,M_\odot$ \citep{Audard10}. The small population of known EXor appears to be dominated by low-mass stars. SPICY 97589 stands out in being particularly young in comparison to other EXors, EX\,Lupi is $\sim 3.5\,\mathrm{Myr}$ compared to $\sim 0.75\,\mathrm{Myr}$ found for SPICY 97589.

The stellar parameters derived here are similar to those derived in \citet{Nagy25}, which was based on optical and NIR spectra and the TiO molecular lines. They fit an M1.0 stellar template with an extinction of $A_V = 3.2\pm0.5$, which they relate to a mass of $0.4\pm0.05\,\mathrm{M_\odot}$. This is not in exact agreement with our values, but the slight difference can be attributed to the different modelling tools employed and the specific evolutionary tracks/isochrones used.

\subsection{YSO Classification}

The estimated age of SPICY 97589 of $\sim 0.75\,\mathrm{Myr}$ likely places this object in the Class I stage of young stars. Although the age is not strictly an indicator of YSO class, it is correlated, and most typical EXor-type objects fall into the Class-II classification. SPICY 97589 does stand out as one of the youngest-age EXors discovered to date. This age stands in contrast to the spectral index, as calculated by \citet{Kuhn21}, of -0.62, which places SPICY 97589 firmly in the Class-II classification. This discrepancy is likely a combination of endemic inaccuracies in isochronal ageing and the potential impact of the strong variability in the initial classification of \citet{Kuhn21}.

Class I objects typically still exhibit an infalling envelope, which can replenish the protoplanetary disk. In the classical view of episodic accretion, objects with an infalling envelope can fuel larger-scale FUor type as the disk can be continuously replenished \citep{Hartmann98,Shu77,Hartmann85}. Later, as the infalling envelope is depleted, the accretion outbursts become smaller in scale and resemble those of EXors. SPICY 97589 appears to deviate from this trend by exhibiting minor outbursts at an early stage; this could indicate different triggering mechanisms.
 
\subsection{Light Curves and Short-Term Evolution}

Looking at the light curves of Gaia23bab in the optical and MIR, they strongly resemble those of other EXor-type objects \citep{Wang23,Kuhn24}. They exhibit a short period of rise and fall, approximately of equal duration, with only a brief period at peak outburst. This differs significantly from FUors, with short and extreme increases in brightness followed by protracted dimming over many years/decades \citep{Clarke05}. The duration of the outbursts in SPICY 97589, 358 days in 2017 and 671 days in 2023, is also characteristic of EXor sources. The outbursts of EX Lup typically last from 60 to 400 days for successive outbursts (of which there have been over 20 since 1945) \citep{Wang23}. V1118 Ori similarly has endured six outbursts in 40 years on monitoring, with the longest being $\sim 2\mathrm{yr}$ in length. SPICY 97589 has now experienced two outbursts in almost 8 years of photometric monitoring; although it is impossible to say if this frequency is periodic or irregular. 

%\textbf{The colour of the outburst needs more research.}

When comparing the pre-outburst SED with the flux-calibrated spectra at different epochs (see Figure\,\ref{fig:SED}), we can monitor the evolution of the outburst. During the outburst, the UVB and VIS regions show greatly enhanced continuum emission, typical of accretion outbursts where accretion luminosity contributes to shorter wavelengths. The NIR excess is only moderate as this region is dominated by warm dust rather than accretion shocks. This is a typical feature of EXor outbursts as seen in EX\,Lup observations \citep{Cruz23}.

Where SPICY 97589 differs from other EXors is in the immediate post-outburst SED. There is a significant drop in flux at all wavelengths (UVB to NIR) compared to pre-outburst; indeed, while the outburst increased the GAIA Gmag by 2.51 mag above quiescence, immediately post-outburst, the Gmag as calculated from the X-Shooter data is $\sim 0.4\,\mathrm{mag}$ below quiescence. This drop is smaller in the UVB and larger throughout the NIR. In the 2022 outburst of EX\,Lup, the immediate post-outburst light curve and spectroscopy returned to the same level as pre-outburst \citet{Cruz23}, triggering the question: why does SPICY 97589 continue to dim post-outburst? We hypothesise that this could be the result of the inner disk being depleted by the outburst, which removes warm dust close to the star and cuts off accretion processes. If true, the post-outburst measured accretion rate may not accurately represent 'normal' quiescence. We may expect the inner disk to be replenished during the long quiescent period preceding the next outburst. Additional epochs of multi-wavelength photometry are required to confirm this.

\subsection{Mass Accretion Rate}
The final calculated mass accretion rate during the outburst is $2.38\pm0.58\times10^{-7}\,\mathrm{M_\odot yr^{-1}}$, and is $4.24\pm0.67\times10^{-9}\,\mathrm{M_\odot yr^{-1}}$ post-outburst; the accretion rate dropped by 2 orders of magnitude. The quiescent accretion rate is largely consistent with the expected accretion rate in Class I/II YSOs, which is usually $\sim 10^{-9}\,\mathrm{to}\,10^{-10}\,\mathrm{M_\odot yr^{-1}}$ \citep{Fiorellino21}. This change in accretion rate is quite dramatic. The 2022 outburst of EX\,Lup saw an increase of 2.5mag in the G-band, and the measured accretion rates at quiescent and outburst were $\sim 4\times10^{-8}$ and $\sim 3\times10^{-7}$, a much more modest change compared to SPICY 97589. This difference can be explained as our post-outburst spectrum does not seem to represent genuine quiescence, but an even dimmer state with lower than expected accretion rates.

Previous studies of SPICY 97589 found broadly similar mass accretion rates during the outburst. Both \citet{Giannini2024} and \citet{Nagy25} find $\dot M \sim 2.0\times10^{-7}\,\mathrm{M_\odot yr^{-1}}$ based on spectra taken $\sim 40$ days and $\sim 30$ days after the peak out the outburst respectively. Both groups employ similar methods, using broadly the same accretion-tracing lines. The similarity in the results from three independent spectra highlights the robustness of the methodology.

Based on the light curve and the measured accretion rates, we can estimate the total mass accreted during the outburst. To do this, we assume that the accretion rate immediately before outburst is the same as immediately after, although this is likely an underestimation as described. We also assume that the change in accretion rate over the outburst follows the same trajectory as the light curve, allowing us to integrate under the curve to obtain the total mass accreted. We find that a total of $\sim 2.3\times10^{-8}\,\mathrm{M_\odot}$ was accreted over 671 days, such a mass would take around 35\,yr to accrete with steady-state, quiescent accretion. This highlights the importance of accretion outbursts on all scales for the assembly of stellar mass.

\subsection{Other Spectral Features} \label{sec:SpectraDis}
\begin{figure}[b!]
    \centering
    \includegraphics[scale=0.5]{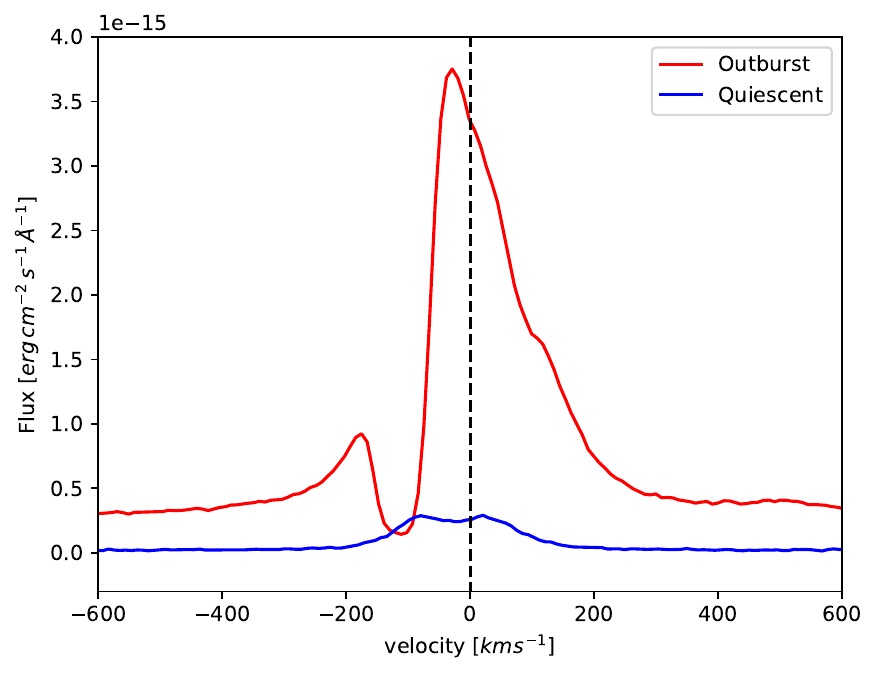}
    \caption{Heliocentric corrected P-Cygni profile of the H$\alpha$ line at a rest wavelength of $656.28\,\mathrm{nm}$ (black dashed line). The blue line is the quiescent spectra taken, while red is the outburst line. The spectra are not continuum-subtracted or offset.}
    \label{fig:BaAlpha}
\end{figure}

The nature of the hydrogen lines differs between the series. The Brackett and Paschen series are in emission only, while the Balmer lines all show P-Cygni profiles as shown in Figure\,\ref{fig:BaAlpha}

This characteristic of blue-shifted absorption and red-shifted emission is found in many YSOs \citep{Lorenzetti09}. It is commonly attributed to outflows such as disk winds, where the outflowing material contributes to the emission while also obscuring the central star, causing absorption \citep{Herbig08}.  Perhaps unusually in the case of SPICY 97589, both the absorption and emission peaks are redshifted; this is only possible in an inclined disk system with a wide opening angle disk wind. In such a system, most of the disk wind flows directly towards the observer, resulting in a redshifted emission peak. The disk wind emitted from the near side of the disk is 'aimed' directly towards the observer while also obscuring the central star, creating a high-velocity red-shifted absorption component. 

In the quiescent spectra, the P-Cygni profile has diminished entirely and is replaced by a weak double-peaked emission. This indicates a 'switching off' of the wind launching mechanism. Instead, it is replaced by a low-velocity, low-angle wind which does not obscure the central star. The rate of wind outflow has long been correlated with the accretion rate in similar cases \citep{Lorenzetti09} such as V1118\,Ori \citep{Herbig08} where a P-Cygni profile during outburst became a weak double-peaked emission during quiescence. 

In addition to Hydrogen lines, a wide range of metallic lines are identified in the spectra: Fe, Na, Si, Mg, Ai, C, and O are all present in the outburst in emission. Following the outburst, the nature of many of these lines changes; some fade to very low or undetectable emission, while others transition into absorption. The Fe II line, a known shock-excited line at 1.644\,$\mu m$, was detected in the outburst spectrum,  but is strongly blended with Br12. 

\subsection{EXor spectral classification}
A key indicator of accretion outbursts is the presence of many strong emission lines and a lack of absorption lines. \citet{Fischer23} characterise EXor spectra as containing strong emission lines in Hydrogen, CO, Fe, Mg and Ca, with Si and Ti lines also. Our outburst epoch spectra, as shown in Figure\,\ref{fig:AllSpectra}, clearly show the presence of all lines in powerful emission, in particualr the hydrogen lines as discussed. A list of the strongest spectral lines and their profiles is provided in Appendix\,\ref{AppB}.  Overall, the spectra strongly resemble other outbursting EXor sources such as EX\,Lup, V118\,Ori and V1143\,Ori \citep{Kospal11,Fischer23,Kuhn24,Herbig08}, and we are confident of identifying SPICY 97589 as a member. One of the most prominent and classical features observed in EXor spectra is the variation in CO bandheads around $2.3-2.4\,\mu m$. During outburst phases, these features appear in emission, whereas during quiescence they become muted absorption features. In SPICY 97589, we observe a strong emission during outburst, followed by fading into very low-level absorption post-outburst.

%\begin{figure*}
    %\centering
    %\includegraphics[scale=0.55]{All_Spectra_Quiet.pdf}
    %\caption{X-Shooter post-outburst spectra across the three instrument arms. The three panels from top to bottom show the spectra from %the UVB, VIS and NIR arms. For reference, the main lines identified in the outburst spectra are highlighted as dashed lines. }
%    \label{fig:AllSpectra}
%\end{figure*}

\section{Conclusions} \label{sec:Conclusion}

Our spectroscopic characterisation has characterised the YSO SPICY 97589, adding it to an exclusive but growing list of accretion outbursting objects.  We summarise our conclusions as follows:
    \begin{itemize}

    \item SPICY 97589 light curves in the optical and MIR show two characteristic $>2\,\mathrm{mag}$ outbursts in 2017 and 2023, of scale, shape and duration similar to many other EXor type outbursts.

    \item The UVB-VIS-NIR outburst spectrum shows a forest of emission lines, as the outbursting inner disk and accretion shocks outshine the stellar photosphere. These lines indicate the presence of a strong disk wind, which is enhanced during the accretion event. Such a spectrum shows strong expected similarities with other EXor outbursts.

    \item We measure the accretion rate using multiple accretion tracing lines to be $\dot M = 2.38\pm0.58\times10^{-7}\,\mathrm{M_\odot yr^{-1}}$. This is $\sim 2$ orders of magnitude above the quiescent accretion rate of $\dot M = 4.24\pm0.67\times10^{-9}\,\mathrm{M_\odot yr^{-1}}$.

    \item We determine the central star to be a 3410\,K object of M3.0 spectral type with a mass and age of $\sim 0.29\pm\,\mathrm{M_\odot}$ and $\sim 0.75\,\mathrm{Myr}$ respectively. This is in line with other observations of other EXor and FUor stars as young, low mass objects. 

    \end{itemize}

\begin{acknowledgements}
    Based on observations collected at the European Organisation for Astronomical Research in the Southern Hemisphere under ESO programmes 112.26Z7.001 and 113.28GX.001.\\
    
    This publication makes use of VOSA, developed under the Spanish Virtual Observatory (https://svo.cab.inta-csic.es) project funded by MCIN/AEI/10.13039/501100011033/ through grant PID2020-112949GB-I00. VOSA has been partially updated by using funding from the European Union's Horizon 2020 Research and Innovation Programme, under Grant Agreement nº 776403 (EXOPLANETS-A)

    We acknowledge ESA Gaia, DPAC and the Photometric Science Alerts Team (http://gsaweb.ast.cam.ac.uk/alerts).
\end{acknowledgements}

\bibliographystyle{aa}
\bibliography{REF}

\appendix

\section{Photometric Sources}\label{AppA}
\begin{table}[h]
    \caption{List of photometric sources and fluxes used to compile the spectral energy distribution. References: PAN-STARRS \citet{Chambers16}, 2MASS \citet{Cutri03}, WISE \citet{Cutri14}.}
    
    \centering
    \begin{tabular}{c c c c} 
        \hline
        \noalign{\smallskip}
        $\lambda$  & Flux  & Error & Source  \\ [0.5ex]
         [$\AA$] &  [$erg s^{-1} cm^{-2} \AA^{-1}$] &  &  \\ 
        \hline
        \noalign{\smallskip}
        4810 & 6.93E-18 & 4.90E-19 & PAN-STARRS\\
        6155 & 2.84E-17 & 2.68E-18 & PAN-STARRS\\
        7503 & 1.25E-16 & 7.95E-19 & PAN-STARRS\\
        8668 & 2.73-16 & 9.35E-18 & PAN-STARRS\\
        9613 & 4.38E-16 & 8.11E-18 & PAN-STARRS\\
        12350 & 7.53E-16 & 3.19E-17 & 2MASS\\
        16620 & 8.02E-16 & 2.88E-17 & 2MASS\\
        21590 & 5.59E-16 & 1.95E-17 & 2MASS\\
        33526 & 2.79E-16 & 7.99E-18 & WISE\\
        46028 & 1.76E-16 & 3.89E-18 & WISE\\
        115608 & 4.00E-17 & 1.99E-18 & WISE\\
        220883 & 1.62E-17 & 9.43E-19 &  WISE\\
        \hline
    \end{tabular}
    \end{table}

\break

\section{Emission Lines}\label{AppB}

\begin{table*}[h]
\centering
\begin{threeparttable}
    \caption{List of Hydrogen lines present in the X-Shooter outburst spectrum and the line characteristics. $\lambda$ is the central wavelength of the line given in air. EW is the equivalent width of the line measured on the outburst spectrum. A dash represents a non-detection of the line in the quiescent spectrum. }
    \centering
    \begin{tabular}{c c c c c} 
        \hline
        \noalign{\smallskip}
        Species & $\lambda$ [nm] & EW [$\AA$] & Profile (Outburst) & Profile (Quiescence)   \\ [0.5ex]
        \hline
        \noalign{\smallskip}
        H$\mathrm{\alpha}$ & 656 & -54.56* & P Cyngi & Double-Peaked Emission \\
        H$\mathrm{\beta}$ & 486 & -9.11* & P Cyngi & Emission \\
        H$\mathrm{\gamma}$ & 434 & -1.19* & P Cyngi & - \\
        \noalign{\smallskip}
        \hline
        \noalign{\smallskip}
        Pa$\mathrm{15}$ & 844 & -22.13 & Emission & Emission\\
        Pa$\mathrm{14}$ & 847 & -2.59 & Emission & - \\
        Pa$\mathrm{13}$ & 866 & -2.63 & Emission & Emission \\
        Pa$\mathrm{12}$ & 875 & -3.28 & Emission & - \\
        Pa$\mathrm{11}$ & 886 & -6.13& Emission & - \\
        Pa$\mathrm{10}$ & 901 & -4.32 & Emission & - \\
        Pa$\mathrm{9}$ & 923 & -5.90 & Emission & Emission \\
        Pa$\mathrm{8}$ & 954 & -6.27 & Emission & Emission \\
        Pa$\mathrm{\delta}$ & 1005 & -7.65 & Emission & Emission\\
        Pa$\mathrm{\gamma}$ & 1094 & -8.74 & Emission & Emission \\ 
        Pa$\mathrm{\beta}$ & 1282 & -22.13 & Emission & Emission \\
        \noalign{\smallskip}
        \hline
        \noalign{\smallskip}
        Br$\mathrm{18}$ & 1534 & -0.72 & Emission & - \\
        Br$\mathrm{17}$ & 1544 & -0.83 & Emission & - \\
        Br$\mathrm{16}$ & 1556 & -1.39 & Emission & - \\
        Br$\mathrm{14}$ & 1589 & -2.10 & Emission & - \\
        Br$\mathrm{13}$ & 1611 & -3.30 & Emission & - \\
        Br$\mathrm{12}$ & 1640 & -5.32** & Emission & - \\
        Br$\mathrm{11}$ & 1681 & -4.30 & Emission & - \\
        Br$\mathrm{10}$ & 1737 & -4.35 & Emission & - \\
        Br$\mathrm{9}$ & 1944 & -7.20 & Emission & - \\
        Br$\mathrm{\gamma}$ & 2166 & -15.79 & Emission \\ 
        \hline
    \end{tabular}
    \begin{tablenotes}
      \small
      \item * Measurements are strongly impacted by p-cyngi profiles in the line; as such, they are not considered when computing the final accretion rate. ** Line is strongly blended with FeII at 1664\,$\mu m$
    \end{tablenotes}
  \end{threeparttable}
    \end{table*}

\begin{table*}[h]
\centering
\begin{threeparttable}
    \caption{Non-exhaustive List of Helium and metallic lines present in the X-Shooter outburst spectrum and the line characteristics. $\lambda$ is the central wavelength of the line given in air, except NaI, which is given in vacuum. EW is the equivalent width of the line measured on the outburst spectrum. A dash represents a non-detection of the line in the quiescent spectrum. }
    \centering
    \begin{tabular}{c c c c c} 
        \hline
        \noalign{\smallskip}
        Species & $\lambda$ [nm] & EW [$\AA$]& Profile (Outburst) & Profile (Quiescence)   \\ [0.5ex]
        \hline
        \noalign{\smallskip}
        CaI & 1689 & -2.45 & Emission & Emission \\
        CaI & 1951 & -5.16 & Emission & Absorption \\
        CaI & 1978 & -5.78 & Emission & Absorption \\
        CaI & 1987 & * & Emission & - \\
        \noalign{\smallskip}
        \hline
        \noalign{\smallskip}
        CaII & 849 & -37.88 & Emission & Emission \\
        CaII & 854 & -41.78 & Emission & Emission \\
        CaII & 866 & -38.25 & Emission & Emission \\
        CaII & 1184 & -2.94 & Emission & - \\
        CaII & 1195 & -2.81 & Emission & - \\
        \noalign{\smallskip}
        \hline
        \noalign{\smallskip}
        HeI & 588 & -1.56 & Emission & Emission \\
        HeI & 668 & -2.10 & Emission & Emission \\
        HeI & 706 & -0.77 & Emission & Emission \\
        HeI & 1083 & - & P-Cygni & P-Cygni \\ %-8.22
        \noalign{\smallskip}
        \hline
        \noalign{\smallskip}
        OI & 777 & -3.14 & Emission & - \\
        OI & 845 & -3.78 & Emission & Emission \\
        OI & 1129 & -3.64 & Emission & - \\
        \noalign{\smallskip}
        \hline
        \noalign{\smallskip}
        MgI & 1183 & -3.50 & Emission & Absorption \\
        MgI & 1209 & -2.10 & Emission & - \\
        MgI & 1488 & -3.25 & Emission & Absorption \\
        MgI & 1503 & -6.7 & Emission & Absorption \\
        MgI & 1504 & * & Emission & Absorption \\ %* -8.91 
        MgI & 1505 & * & Emission & - \\
        MgI & 1577 & -4.51 & Emission & - \\
        MgI & 1711 & -3.71 & Emission & Abosrption \\
        \noalign{\smallskip}
        \hline
        \noalign{\smallskip}
        NaI & 2206 & -1.68 & Emission & Absorption \\
        NaI & 2209 & -1.27 & Emission & Absorption \\
        \noalign{\smallskip}
        \hline
        \noalign{\smallskip}
        FeI & 1144 & -1.83 & Emission & Absorption \\
        FeI & 1164 & -2.98 & Double-Peaked Emission & Absorption \\
        FeI & 1169 & -2.74 & Double-Peaked Emission & Absorption\\
        FeI & 1179 & -2.38 & Double-Peaked Emission & Absorption \\
        FeI & 1198 & -5.12 & Emission & Absorption \\
        FeI & 1329 & -1.88 & Emission & Absorption \\
        FeI & 1455 & -3.45 & Emission & - \\
        FeI & 1530 & -1.84 & Emission & - \\
        \noalign{\smallskip}
        \hline
        \noalign{\smallskip}
        FeII & 1112 & -0.76 & Emission & Absorption \\
        \noalign{\smallskip}
        \hline
        \noalign{\smallskip}
        SiI & 1199 & * & Emission & - \\
        SiI & 1200 & * & Emission & - \\
        SiI & 1208 & -2.01 & Emission & - \\
        SiI & 1227 & -1.78 & Emission & Absorption \\
        \noalign{\smallskip}
        \hline
        \noalign{\smallskip}
        AlI & 1313 & -2.58 & Double-Peaked Emission & Absorption \\
        AlI & 1315 & -2.50 & Double-Peaked Emission & Absorption \\
        AlI & 1329 & -1.78 & Emission & Absorption \\
        
        \hline
    \end{tabular}
    \begin{tablenotes}
      \small
      \item * Measurements are not reliable due to strong blending of lines
    \end{tablenotes}
  \end{threeparttable}
    \end{table*}

\end{document}